\documentclass[]{spie}  

\usepackage{amsmath,amsfonts,amssymb}
\usepackage{graphicx}
\usepackage[colorlinks=true, allcolors=blue]{hyperref}\usepackage{xspace}
\usepackage{upgreek}
\usepackage{amsmath,amsfonts,amssymb}
\usepackage{graphicx}
\usepackage{color, colortbl} 
\usepackage{booktabs} 
\usepackage{multirow} 
\usepackage[inline]{enumitem} 

\usepackage{wasysym}
\usepackage{physics} 
\usepackage{textcomp,gensymb} 
\usepackage{subcaption} 
\usepackage{siunitx}
\usepackage{wrapfig}

\newcommand{\mytilde}{\raise.17ex\hbox{$\scriptstyle\mathtt{\sim}$}} 
\providecommand{\sorthelp}[1]{} 

\def\3he{$^3{\rm He}$}

\def\lsim{\mathrel{\lower2.5pt\vbox{\lineskip=0pt\baselineskip=0pt
           \hbox{$<$}\hbox{$\sim$}}}}

\def\gsim{\mathrel{\lower2.5pt\vbox{\lineskip=0pt\baselineskip=0pt
           \hbox{$>$}\hbox{$\sim$}}}}

\def\procspie{\ref@jnl{Proc.~SPIE}}   

\title{The Balloon-Borne Large Aperture Submillimeter Telescope Observatory}

\author[a]{Ian Lowe}
\author[b]{Gabriele Coppi}
\author[c]{Peter A. R. Ade}
\author[d]{Peter C. Ashton}
\author[e]{Jason E. Austermann}
\author[e]{James Beall}
\author[f]{Susan Clark}
\author[g]{Erin G. Cox}
\author[a]{Mark J. Devlin}
\author[a]{Simon Dicker}
\author[e]{Bradley J. Dober}
\author[b]{Valentina Fanfani}
\author[h]{Laura M. Fissel}
\author[i]{Nicholas Galitzki}
\author[e]{Jiansong Gao}
\author[j]{Brandon Hensley}
\author[e]{Johannes Hubmayr}
\author[j]{Steven Li}
\author[k]{Zhi-Yun Li}
\author[l]{Nathan P. Lourie}
\author[m]{Peter G. Martin}
\author[n]{Philip Mauskopf}
\author[b]{Federico Nati}
\author[g]{Giles Novak}
\author[c]{Giampaolo Pisano}
\author[j]{L. Javier Romualdez}
\author[n]{Adrian Sinclair}
\author[o]{Juan D. Soler}
\author[c]{Carole Tucker}
\author[e]{Michael Vissers}
\author[e]{Jordan Wheeler}
\author[g]{Paul A. Williams}
\author[b]{Mario Zannoni}

\affil[a]{University of Pennsylvania, 209 South 33rd Street, Philadelphia, PA 19104, USA}
\affil[b]{University of Milano-Bicocca, Piazza della Scienza 3, 20126 Milano (MI), Italy}
\affil[c]{Cardiff University, The Parade, Cardiff CF24 3AA, United Kingdom}
\affil[d]{Lawrence Berkeley National Laboratory, 1 Cyclotron Rd, Berkeley, CA 94720}
\affil[e]{NIST-Boulder, 325 Broadway, Boulder, CO 80305}
\affil[f]{Institute for Advanced Study, 1 Einstein Dr, Princeton, NJ 08540}
\affil[g]{Northwestern University, 1800 Sherman Ave, Evanston, IL 60201}
\affil[h]{Queen's University, 99 University Ave, Kingston, ON K7L 3N6, Canada}
\affil[i]{University of California San Diego, 9500 Gilman Dr, La Jolla, CA 92093}
\affil[j]{Princeton University, Jadwin Hall, Washington Road, Princeton, NJ 08544}
\affil[k]{University of Virginia, 530 McCormick Road, Charlottesville, VA 22904}
\affil[l]{MIT Kavli Institute for Astrophysics and Space Research, 70 Vassar St, Cambridge, MA 02139}
\affil[m]{CITA, University of Toronto, 60 St. George St., Toronto, ON M5S 3H8 Canada}
\affil[n]{Arizona State University, 550 E Tyler Drive, Tempe, AZ 85287}
\affil[o]{Max Planck Institute for Astronomy, Königstuhl 17, 69117 Heidelberg, Germany}

\authorinfo{Further author information: (Send correspondence to Ian Lowe)\\Ian Lowe: E-mail: ianlowe@sas.upenn.edu, Telephone: 1 484 919 6228}

\begin{document} 
\maketitle

\begin{abstract}
    \paragraph{}
    The BLAST Observatory is a proposed superpressure balloon-borne polarimeter designed for a future ultra-long duration balloon campaign from Wanaka, New Zealand. To maximize scientific output while staying within the stringent superpressure weight envelope, BLAST will feature new 1.8m off-axis optical system contained within a lightweight monocoque structure gondola. The payload will incorporate a 300~L $^4$He cryogenic receiver which will cool 8,274 microwave kinetic inductance detectors (MKIDs) to 100mK through the use of an adiabatic demagnetization refrigerator (ADR) in combination with a $^3$He sorption refrigerator all backed by a liquid helium pumped pot operating at 2~K. The detector readout utilizes a new Xilinx RFSOC-based system which will run the next-generation of the BLAST-TNG KIDPy software. With this instrument we aim to answer outstanding questions about dust dynamics as well as provide community access to the polarized submillimeter sky made possible by high-altitude observing unrestricted by atmospheric transmission. The BLAST Observatory is designed for a minimum 31-day flight of which 70$\%$ will be dedicated to observations for BLAST scientific goals and the remaining 30$\%$ will be open to proposals from the wider astronomical community through a shared-risk proposals program.
\end{abstract}
\section{Introduction}
\paragraph{}
The Balloon-borne Large Aperture Submillimeter Telescope - BLAST Observatory is a proposed superpressure balloon (SPB) experiment, designed to provide access to the polarized submillimeter sky. The BLAST collaboration has been building balloon-based instruments to observe the submillimeter thermal emission of dust since 2003. The first LDB flight from Antarctica in 2006 utilized 270 unpolarized detectors at 250, 350, and 500~$\mu m$ to observe high-redshift star formation\cite{enzo_blast}.  In 2009 and 2011, the same instrument was upgraded with polarization capabilities to observe star formation in our own galaxy\cite{tyr_blastpol_spie,fissel_blastpol}.  Finally, in 2020 we flew a completely new experiment with 10 times the detectors which had been upgraded to polarized MKIDs\cite{nate_receiver,nate_instrument}. These experiments have flown a total of five times\cite{enzo_blast, tyr_blastpol_spie} and provided a wealth of technological and scientific advances, from showing that the dominant star-forming period occurred at z$>$1\cite{devlin09} to mapping the magnetic fields within star-forming regions of our own galaxy\cite{Fissel_2019} and demonstrating the viability of kilopixel MKID arrays on stratospheric balloons. 

The BLAST Observatory will address key outstanding unresolved questions in our understanding of star formation, turbulence in the interstellar medium (ISM), and the properties of interstellar dust, achieved through a systematic survey of polarized dust emission near the characteristic dust emission intensity peak of 100 - 500$\mu$m. We will also make the first statistical sample of high-resolution ($<$ 0.1pc) magnetic field maps covering entire molecular cloud complexes, determining the role of magnetic fields in regulating the formation of clouds, filaments, and dense star-forming cores\cite{soler_hennebelle,Fissel_2019,fisselmagneticfields}. Our surveys will provide the critical link between {$Planck$} all-sky polarization maps and ALMA's ultra-high resolution, giving an unprecedented view of how magnetic fields shape the formation of individual stars. Furthermore, these high resolution maps of the diffuse ISM will quantify properties of MHD turbulence, and test turbulence dissipation mechanisms\cite{burkhart09,Falceta_Gon_alves_2014}. Finally, by simultaneously measuring polarized dust emision at three THz frequencies, the BLAST Observatory aims to test models of dust grain composition. In particular we will determine whether one or more distinct varieties of dust exist\cite{Hensley2019}, and characterize the evolution of dust properties from diffuse gas to cold cores, revealing the mechanical, chemical, and thermodynamic processes that shape interstellar grains.

The BLAST Observatory will take advantage of lessons learned from decades of millimeter and submillimeter balloon observations as well as recent technological advances. Previous flights utilized the mature circumpolar flight (from Kiruna or McMurdo) LDB campaigns, providing access to the limited near-polar sky. Access to a statistically significant sample of star forming regions requires the large sky coverage enabled by a mid-lattitude flight from the Wanaka base in New Zealand providing is critical to the mission’s success. The 1.8 meter low-emission, off-axis Gregorian telescope feeds a 100~mK cryogenic camera with three focal plane arrays with a total of 8,274 polarization-sensitive detectors at 175, 250, and 350~$\mu$m.  Previous advances in detector and readout technology\cite{sam_roaches, sam_ltd} will be incorporated to vastly improve the mapping speed over earlier instruments. 

The BLAST collaboration pioneered the concept that balloon missions should be observatories available to the entire community, with BLAST-TNG offering 25$\%$ of the time for external proposals. After minimal advertising of the shared-risk program, the BLAST team received proposals for over 300$\%$ of the allotted time from community members all over the globe. In light of this, the BLAST Observatory will raise the total time set aside for the shared-risk portion to 30$\%$ of the flight, which will be more promoted more heavily and earlier on to maximize availability to the community.

In this paper we detail the mechanical, optical, and cryogenic design of the new BLAST Observatory instrument as well as the scientific goals. Section \ref{sec:Gondola} describes the design of the new super-pressure envelope gondola. Section \ref{sec:optics} discusses the redesigned off-axis optical system and how this interplays with the receiver. Section \ref{sec:cryo} details the design of the cryogenic receiver as well as the new 100mK cryogenic system we plan to implement. The next generation readout system and MKIDs are presented in Section \ref{sec:detectors}. Finally, the scientific goals of the project are described in Section \ref{sec:science}.

\section{Gondola}
\label{sec:Gondola}
\subsection{Mechanical structure}
\begin{figure}
    \centering
    \includegraphics[scale = 0.25]{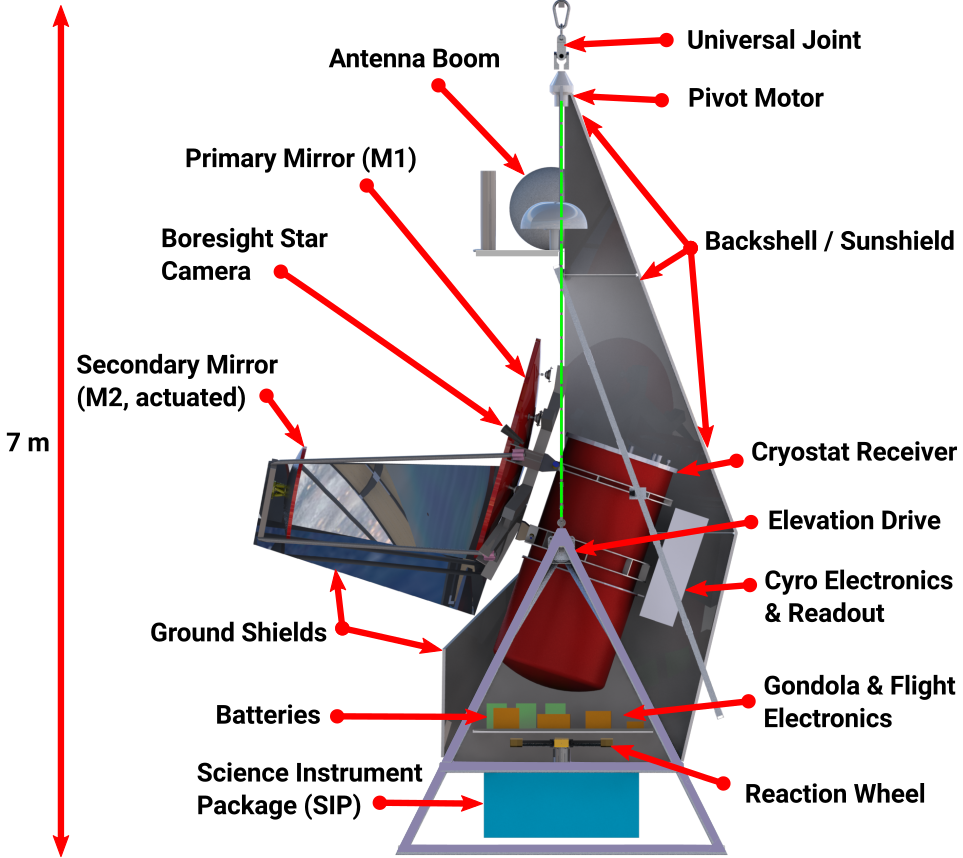}
    \caption{Cutaway render of the BLAST Observatory gondola with critical payload components labeled.}
    \label{fig:gondola}
\end{figure}

\paragraph{}
The BLAST Observatory gondola has been designed to fit within the weight envelope of the superpressure balloon payload while maintaining all performance specifications required for our scientific goals. The gondola consists of two distinct sections, the outer and inner frames, with an interface at the elevation axis. A full mockup of the gondola with component locations labeled and shown can be found in Figure \ref{fig:gondola}. The inner frame is the aluminum structure which houses all of the optical elements of the payload as well as the scientific readout electronics. The inner frame provides a rigid mounting place for the cryogenic receiver as well as the telescope which ensures that the alignment between the two does not drift appreciably as the elevation is changed. In addition, the detector readout electronics are attached to the rear of the inner frame to provide counterbalance weight as well as minimize the path length for the coaxial cables. The outer frame combines an aluminum truss structure with a monocoque shell to create a stiff structure for elevation control. The truss acts as a cradle for the instrument while sitting on the ground as well as an attachment point for the remainder of the electronics, batteries, and CSBF equipment. The shell is made up of honeycomb aluminum panels and is coupled to the suspension cables to provide stress relief during launch and define the suspended geometry of the payload. These panels also act as natural Sun shields for the instrument and provide a rigid mounting place for the solar panels.

\subsection{Pointing control and sensing}
\paragraph{}
Pointing for the BLAST gondola is achieved through an elevation drive and a azimuth control system. The elevation drive is located at the interface of the inner and outer frames and consists of a geared-down harmonic drive (CSF-65-160-GH) driven by a high-torque stepper motor. Cryogen boiloff over the course of the flight causes the inner frame to become unbalanced and require increasing torque to control the pointing. This has been overcome in previous rotary-drive based elevation control through the use of a balancing system which drives a weight to offset the lost inertia. By switching to the new system we retain the necessary pointing resolution through the use of the stepper motor while the harmonic drive provides the requisite torque stiffness to control elevation pointing even under more than twice the maximum modeled imbalance. Protection for these drives during launch and termination is provided through a stepper motor driven pin locking mechanism. The azimuth control system achieves pointing through a combination of a flywheel located below the main payload and a pivot motor located just below the connection to the balloon. The flywheel acts as a fine azimuth pointing system through the control of the angular momentum of the gondola by spinning up or down. Coarse pointing must be achieved through the use of the pivot, which torques against the flight train to transfer angular momentum into the balloon and prevents the saturation of the flywheel pointing capabilities. This combination of azimuth pointing systems has been shown to provide stability and scanning at the 1-arcsecond level as well as provide pendulation damping and roll stability\cite{Romualdez2020}.

While the motors control the physical pointing of the gondola, the pointing determination is achieved through a suite of sensors which are combined into an in-flight pointing solution. For coarse determination, we have a magnetometer, tilt sensors on the gondola, an elevation encoder, a quad-GPS system, and an array of Sun sensors, all of which provide low relative noise but less precise measurements of the gondola attitude. We also utilize much more precise but higher relative noise fiber optic gyroscopes (KVH DSP-1760) which provide the drift of the gondola. The BLAST Observatory attitude determination software takes into account all of this data and the associated sensor quality to determine the in-flight pointing solution. This solution is updated by the absolute pointing sensors, which includes the three integrating star cameras that return celestial pointing from images of the sky. These cameras provide precise pointing determination at a rate of $\sim$0.1~Hz. Together, these systems are capable of providing better than 2$\arcsec$ RMS pointing accuracy during flight.

\section{Optics}
\label{sec:optics}
\begin{figure}[h]
    \centering
    \includegraphics[scale = 0.25]{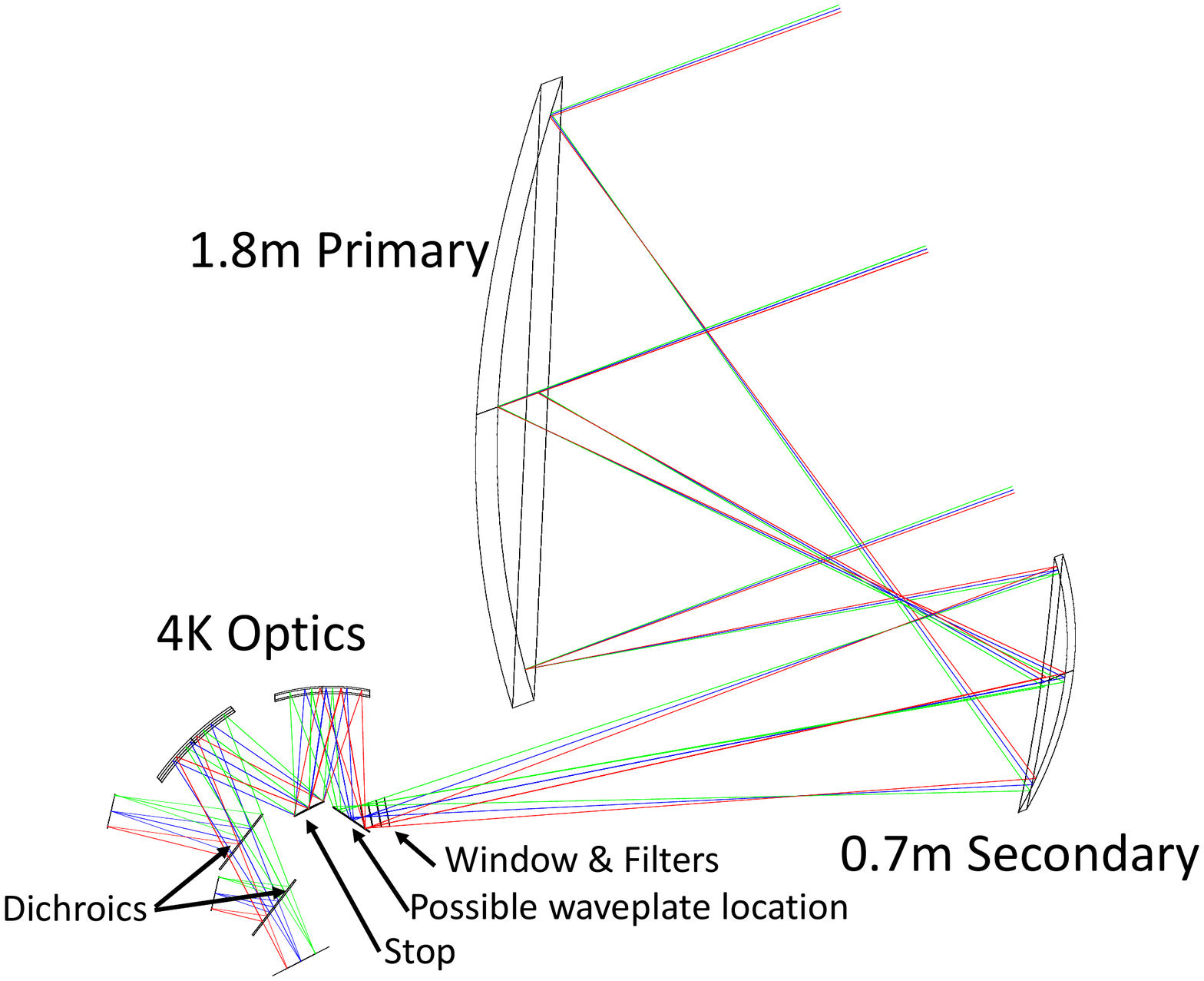}
    \includegraphics[scale=0.25,angle=90]{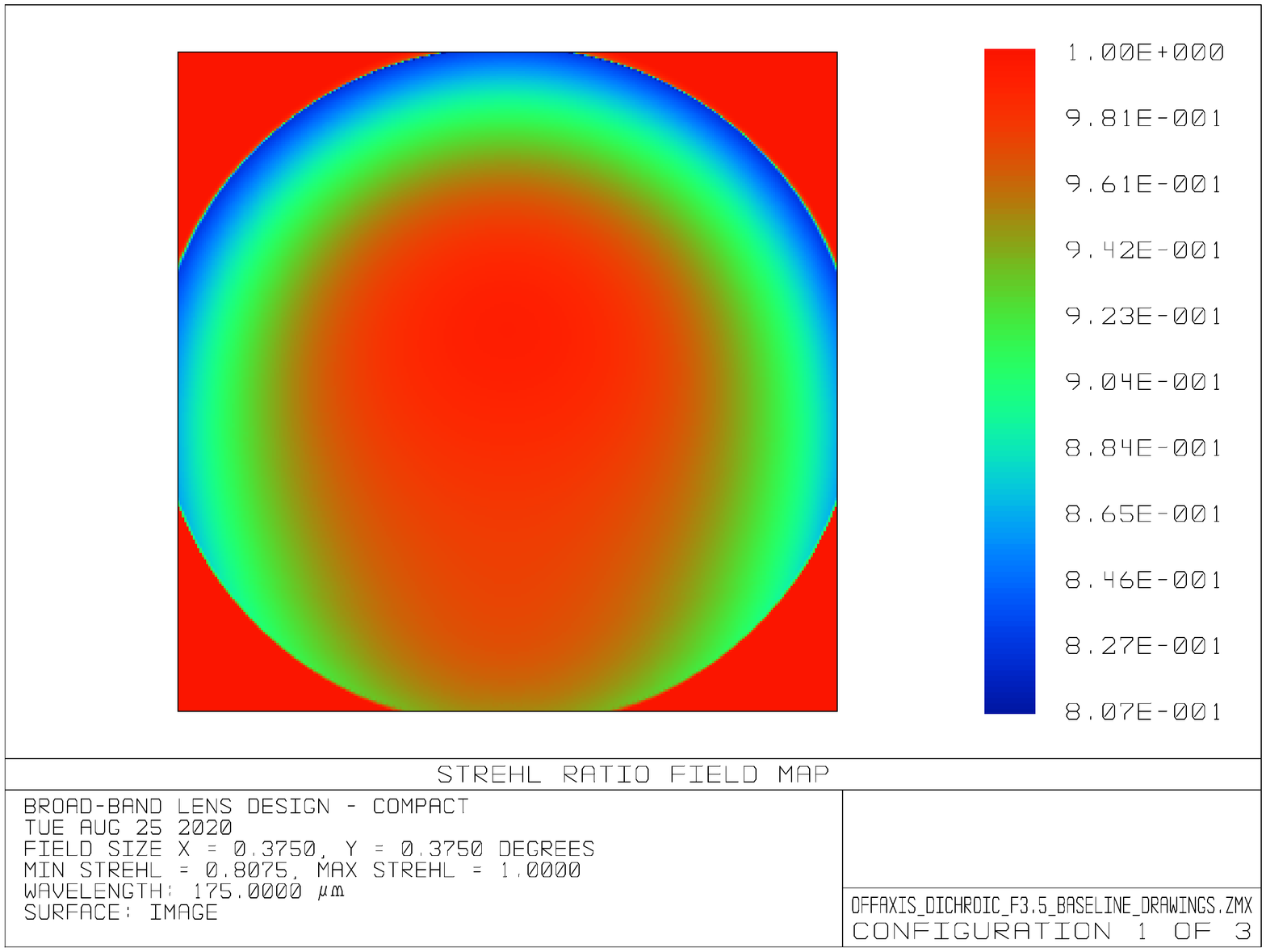}
    \caption{(left) Ray-tracing for design of the BLAST Observatory optics showing the external optics as well as the reimaging optics internal to the cryogenic receiver. The arrays lie in three-dimensional space which has been suppressed for clarity. (right) Map of the Strehl ratio showing diffraction limited performance across the 0.75~$\degree$ focal plane at 175$\mu m$. Simulations at 250 and 350$\mu m$ predict increased performance.}
    \label{fig:optics_design}
\end{figure}

\paragraph{}
The BLAST Observatory has changed the warm optical system from an on-axis telescope to an off-axis one. The new design is a modified Gregorian off-axis telescope in which the primary and secondary mirrors are both conic sections and slightly tilted. This produces the same image quality as the equivalent on-axis design while providing key advantages such as reduced in-band emission and elimination of diffraction off the secondary support structure. This system results in diffraction limited performance over a more than 1~$\degree$ field of view (fov) at 175$\mu m$, our highest frequency band. A full rendering and the modeled performance at 175$\mu m$ can be found in Figure \ref{fig:optics_design}. 

The cryogenic optics remain much the same as previous BLAST iterations, forming an asymmetric Offner relay configuration with two curved mirrors and a curved stop. This configuration produces an f-number of 5 which matches the optics to our desired array fov of 0.75\degree. Additionally, we use the same style flat, telecentric focal plane arrays which remain diffraction limited across the entire array. We achieve simultaneous mapping of the entire fov in all three bands through the use of dichroic beam splitters which split the incident light into three beams towards the appropriate array. In lieu of a transmissive half-wave plate, we opt for a flat mirror near the vacuum window which can be replaced by a spinning reflective half-wave plate should the detector noise characteristics require it. In addition to the change in dichroic beamsplitters, the filter stack will be largely redesigned to ensure minimized loading on the 100mK stage while maximizing optical efficiency in-band. 

\section{Cryogenic receiver}
\label{sec:cryo}
\begin{wrapfigure}{L}{0.5\textwidth}
    \centering
    \includegraphics[scale=0.39]{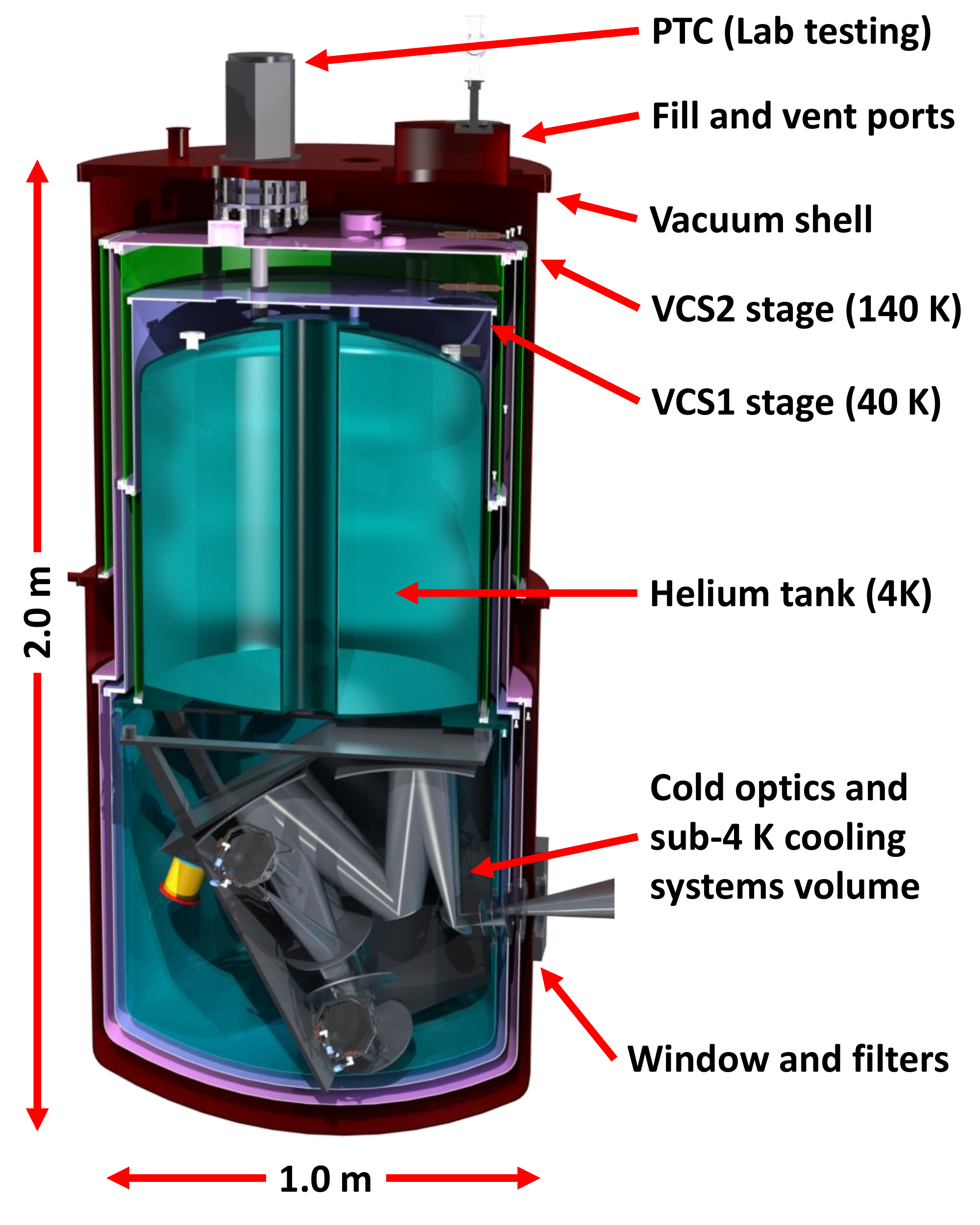}
    \caption{The BLAST Observatory cryogenic receiver. The 300~L helium tank vents through the two heat exchangers located on the vapor-cooled shields to minimize thermal loading on the coldest stage for a predicted 31-day minimum flight time. Additionally, the receiver now features a pulse-tube cooler interfacing directly with the 4~K stage for ground testing with minimal helium use.}
    \label{fig:cryostat}
\end{wrapfigure}

\paragraph{}
The BLAST Observatory will feature a cryogenic receiver with a design similar to that flown on the BLAST-TNG experiment\cite{lourie2018}. A labeled cross-sectional rendering of the new receiver is displayed in figure \ref{fig:cryostat}. The receiver consists of a series of concentric vapor-cooled shields (VCS) which are coupled to the boiloff of the liquid helium reservoir through a pair of heat exchangers on the vent pathway. These VCS equilibrate at roughly 140 and 40~K, which provides a significant reduction in the radiative and conductive loading onto the 4~K helium bath. To further reduce the loading, each stage is wrapped in superinsulation layers which minimizes the radiative coupling between shells. Conductive loading between each stage is minimized by using long and thin G10 offsets as the mechanical connections between each shell. The innermost stage sits at 4~K and consists of the newly enlarged helium tank on one side and the cold plate/optics box on the other side. The helium tank has been redesigned to feature a new 3-port system consisting of a dedicated liquid fill port, heat exchanging boiloff vent, and emergency relief valve. Reconfiguring the liquid tank in this manner increases the operational safety while simplifying fill operations and reducing the likelihood of damaging the heat exchangers. 

The detector array cooling system for the BLAST Observatory cryostat has also been redesigned and now features three stages from 1.8K to 100~mK. The first stage is the pumped pot which is designed to provide 20+~mW of cooling power at 1.8~K through the evaporation of liquid helium at 5-10~mBar. 
This stage provides both a thermal intercept of the 4K parasitic load as well as a thermal sink for the cycling of the two colder stages. We expect this stage to remain largely the same as that used in BLAST-TNG with minor reconfigurations to interface within the new optics and cooling geometry. The second stage is the $^3$He sorption fridge which was previously used to cool BLAST arrays to 275~mK and will act as a secondary thermal intercept for the coldest stage. The final stage is a new adiabatic demagnetization refrigerator (ADR) which will cool the detector arrays to 100~mK, reducing the thermal recombination noise of the MKIDs, and opening up the possibilities of new detector materials such as low transition temperature TiN and Al, both of which feature significantly lowered $\frac{1}{f}$ noise knees.

The redesign of the cryostat also features a new interface for a removable pulse tube cooler (PTC) for laboratory testing. The PTC cold head will attach directly to the helium tank while the intermediate stage will cool both VCS through a combined attachment point. By using a PTC for laboratory testing we will limit the liquid helium use to the pumped pot system which will greatly reduce the required quantity for testing. This change will allow us to operate within the constraints of liquid helium availability given the current and predicted shortages.

\section{Detectors and Readout}
\label{sec:detectors}
\begin{figure}[t]
    \centering
    \includegraphics[scale=0.35]{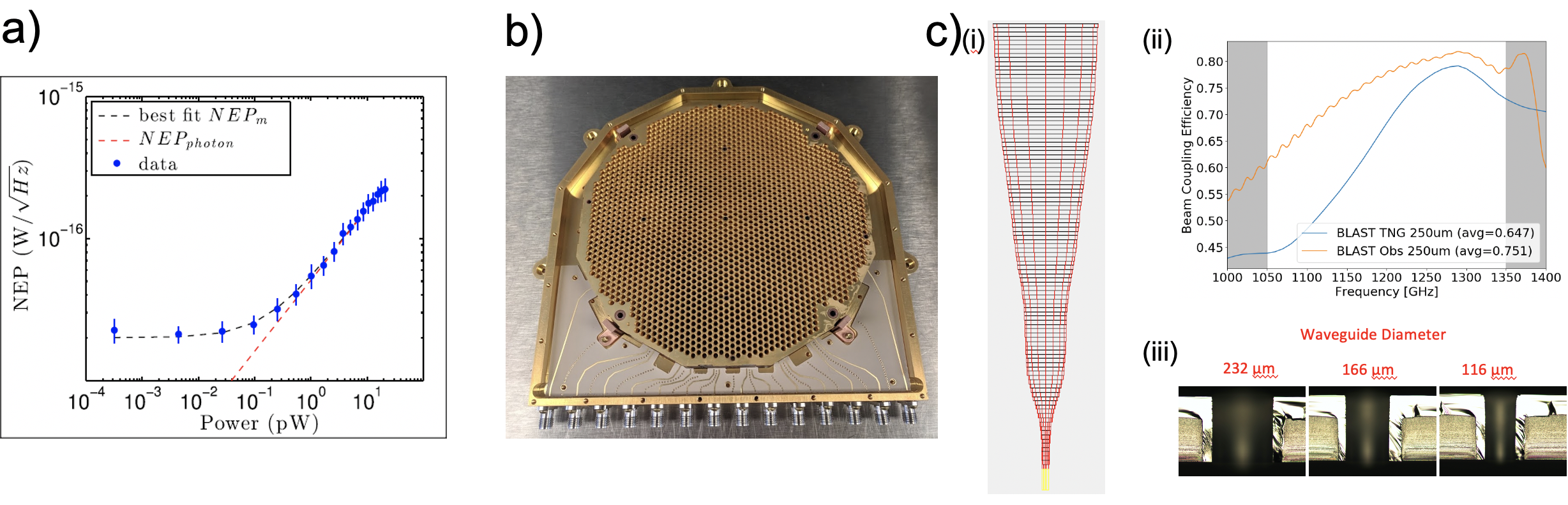}
    \caption{a) Detector NEP versus incident optical power curve for the BLAST-TNG MKIDs showing the absolute noise floor as well as the transition to the linear with loading noise regime. b) TolTEC 150~mm MKID array. the BLAST Observatory will use this same size with a modified mounting scheme to vastly increase the number of detectors per array. c) (i) Preliminary design for the 250~$\mu m$ silicon platelet feedhorns (ii) Expected coupling performance of the platelet feedhorns versus performance of the monolithic metal feedhorn array at 250~$\mu m$ (iii) Test fabrication of platelets with appropriate sizing for the BLAST Observatory.}
    \label{fig:det}
\end{figure}

The BLAST Observatory will fly a total of 8,274 polarized MKIDs spread across three bands at 175, 250 and 350~$\mu m$. The baseline design of the detectors will be the same TiN resonators as were flown on the BLAST-TNG\cite{sam_ltd} experiment while taking advantage of the recent developments in wafer size capabilities for the TolTEC experiment to vastly increase the number of detectors per array\cite{austermann18}. These detectors have excellent polarization response and resilience to cosmic ray strikes in flight. We will also benefit from increase detector yield (BLAST-TNG 70-85$\%$) through the post fabrication resonator editing techniques that have been pioneered in the last few years\cite{liu17}. Our detectors will be optically coupled to the beams through new layered silicon-platelet feedhorns\cite{simon16} which both reduces the microphonics and retains precision alignment at cryogenic temperatures through an all silicon focal plane unit. The platelet feedhorns also have the advantage of high precision and the capability of producing simulation-optimized profiles (Figure \ref{fig:det}c) which predict a 16$\%$ increase in the beam-coupling efficiency at 250~$\mu m$ as compared to the equivalent BLAST-TNG metallic feedhorns.

In addition to the baseline detectors, we plan to work on the development of ``drop in" replacement detector technologies. These other options have the potential to significantly reduce the low frequency noise knee of the detectors, further increasing the quantity and quality of the science we will be able to produce. For these drop-in replacements we are targeting the emerging technologies of the low transition temperature TiN as well as Al MKIDs, both of which are showing promise in the lab as the next generation of detector composition. Both of these options require a 100~mK base temperature being planned for the BLAST Observatory. As discussed in Section \ref{sec:cryo}, we have reconfigured our baseline cryogenic configuration to meet this 100~mK condition and take advantage of these upgrades.

In addition to a revamp of the detectors, the BLAST Observatory readout will be fully rebuilt for the upcoming payload. BLAST-TNG flew a ROACH2-based system which consumed 300~W of power and had a mass of $\sim$250~lbs, far beyond what is feasible for a ULDB mission. Additionally, the old readout system would not be capable of reading out the planned 1600+ detectors per readout channel that the BLAST Observatory is targeting. In order to match our readout to the power and weight budget as well as the minimum readout density target, we have begun development on a system based off of the Xilinx RFSoC ZCU111 board. The firmware architecture will be based largely off the BLAST-TNG readout, however the readout bandwidth will be increased from 512~MHz to 4~GHz and the maximum number of detectors from 1024 to 2048. On top of this major increase in bandwidth and multiplexing capability, each board contains 4 of these separate readout chains with a total power draw of only 30~W. With the current detector configuration, the BLAST Observatory will nearly triple the total detector count while reducing power draw by more than 80$\%$. The increase in readout electronic capabilities and onboard hardware will also allow the BLAST Observatory team to implement more sophisticated elements of the software that were too computationally expensive for BLAST-TNG such as tone tracking algorithms that will further enhance the instrument sensitivity and dynamic range of accessible signals.
\section{Scientific Goals}
\label{sec:science}
\subsection{How do magnetic fields affect star formation}
\paragraph{}
The question of how stars form is crucial to many areas of astrophysics, from the evolution of Galactic structure to planet formation. Coming up with a fully consistent model of star formation is challenging because of the wide range of scales over which star formation takes place: from Giant Molecular Cloud (GMC) scales ($10^{21}$~cm) down to the scale of individual stars ($10^{11}$~cm). In addition, this theory must be able to explain why the stellar formation rate is approximately 1$\%$ of that predicted through free-fall collapse\cite{krumholz14}. This low observed star formation efficiency is most likely due to regulation from turbulence, magnetic fields, and young star feedback\cite{mckee_ostriker,padoan14,burkhart2018,krumholz2019}, but the relative importance of these different forms of regulation is still a subject of active debate. A key challenge is lack of observations, particularly the difficulty of making detailed measurements of magnetic fields across a large sample of molecular clouds. Using the tendancy of dust grains to align with their long axes perpendicular to their local magnetic field, we can use the linearly polarization thermal dust emission to map the cloud magnetic field projected on the sky.

We seek to measure the orientation and strength of the magnetic field through observations of the polarized emission of the dust in at least 100 clouds, as well as synthetic polarization of simulated molecular clouds from a variety of viewing angles so that we can directly compare theory and observations. We will apply statistical methodologies such as \emph{pNS relations}\cite{fissel16},\emph{ Polarization Dispersion Analysis (PDA)}\cite{davis51,chandrasekhar_fermi}, and\emph{ Histograms of Relative Orientations (HRO)}\cite{soler13} to probe the properties of the magnetic fields within these clouds. These data will also be combined with data from \textit{Planck} and ALMA to understand the role of magnetic fields across all stages of star formation: from the formation of molecular clouds from the diffuse ISM, to the formation of dense cloud substructures such as filaments, and the collapse of individual dense molecular cores to form individual stellar systems. With this vast amount of data we will be able to quantify the energy density ratio between magnetic fields, cloud self-gravity, and turbulence and look for correlations with cloud star formation efficiency.

\subsection{Quantifying the behavior of the turbulent interstellar medium}
\paragraph{}
Turbulence not only plays an important role in the formation of stars, but also in the behavior of the diffuse interstellar medium. This magnetohydrodynamic (MHD) turbulence is theorized to produce correlated and scale-dependent structures in the density, velocity, and magnetic fields within the ISM\cite{Elmegreen:2004}. Of specific interest are the \emph{polarized} power spectra of the thermal dust emission, as these will provide new information about the scales at which events such as energy injection and dissipation occur within the magnetic ISM. Turbulence theories predict that the power spectrum should flatten at large angular scale (low multipole $\ell$) where Galactic scale processes, supernovae, and winds dump energy into the ISM and, on small angular scales (high $\ell$), we should expect to see a drop off in the power with a knee corresponding to the scales at which the energy dissipates. The \emph{Planck} experiment measured the dust power spectra\cite{Caldwell:2017,Kritsuk:2018,Kim:2019} in the diffuse ISM. Their analysis, however, was restricted to an $\ell$ range from 60-400\cite{Planck2018XI} due to their angular resolution and sensitivity. The BLAST Observatory will measure the power spectrum of targeted regions of sky out to small angular scales. These high dynamic range measurements of the diffuse ISM polarized power spectrum will be necessary to further develop this field. 

While the flattening at low $\ell$ has also not been robustly measured, the BLAST Observatory mission will focus on constraining the polarized power spectrum at high $\ell$. With our constraining power out to a $\ell \sim 2*10^4$ we seek to understand the mechanism behind the dissipation of energy into the diffuse ISM. Our measurements of the diffuse ISM will produce \emph{polarized} auto- and cross-power spectra will either determine or place strong constraints on the scales associated with energetic dissipation in the diffuse ISM.

\subsection{Determining the composition of interstellar dust}
\paragraph{}

Our final scientific objective is to understand the composition and properties of the interstellar dust itself. In addition to providing a vehicle for investigating the structure of the magnetic fields, interstellar medium density, and velocity, the dust particles in the ISM provide a surface for chemical synthesis and eventual star and planet formation. Decades of research into these interstellar dust grains have left questions about the composition and population of dust species. Dust models predict the formation of two distinct types of dust, the ``carbonaceous" and ``silicate" dust grains, both of which have distinct behaviors. The carbonaceous grains are expected to equilibrate at higher temperatures and exhibit less polarized emission than their silicate counterparts\cite{Hoang2016, Chiar+etal_2006}. Results from the BLASTPol and \emph{Planck} experiments have shown that theories predicting the fractional polarization or the SED differences between intensity and polarization were unable to reproduce the measured values, forcing further theoretical developments. 

Recent theoretical advances have sought to explain this behavior in the far-infrared region of the spectrum. One approach is to assume highly aspherical grains producing the high polarization fractions with the two populations characterized by similar opacity laws producing the long-wavelength behavior seen in the CMB datasets\cite{Guillet2018}. The other approach has been to model the dust as a single component with the difference in polarization coming down to the grain surface chemistry\cite{Hensley2020}. Distinguishing between these two theoretical models has been difficult as the the behavior in the CMB survey wavelengths is in the Rayleigh-Jeans tail, scaling as \emph{T}, which is completely degenerate with the column density in the sightline. The BLAST Observatory, with bands centered at 175, 250, and 350$\mu m$, is uniquely suited to measure the polarized dust emission near the SED peak, where the two models differ significantly in their predictions. Additionally, distinguishing between these two models requires observations in regions where the expected grain temperature is primarily a function of composition, rather than the local radiation field. With our measurements of the polarized emission of the diffuse ISM we will be able to robustly determine the high-frequency emission behavior of these dust grains, providing evidence to distinguish between competing models.

\section{Conclusions}
\label{sec:conc}
\paragraph{}
The BLAST Observatory is a proposed superpressure balloon experiment that combines the development of emerging technologies with science at the forefront of far-infrared astrophysics. We are building the next-generation of MKID readout which will more than quadruple the available bandwidth for the detectors, allowing for larger resonator spacing in fabrication to reduce collisions, while simultaneously doubling the number that can be read out over a single coaxial line. We will vastly increase the number of detectors from BLAST-TNG to 8,274 through the use of new wafer production methods to produce much larger focal planes as well as work to develop new detector technologies with the potential for large gains in the low frequency noise regime. To enable all of this our cryogenic system has been updated to feature a new ADR with cooling capabilities down to 100mK. To maximize the scientific output of the mission, we have fully redesigned our optics system as an off-axis telescope, removing the diffraction, emission, and signal blockage from the secondary mirror supports for a significant gain in mapping speed. Additionally, we have shifted our wavebands to 175, 250, and 350$\mu m$ to focus on measuring emission closer to the dust SED peak. With a 30+ day mid-latitude flight we will aim our polarized MKIDs at regions of the diffuse ISM, map specific GMCs of great interest, and survey hundreds of GMCs near the Galactic plane to provide a statistical sample of their overall characteristics. Our measurements will provide answers about the composition of the interstellar dust, the properties of the turbulence in the ISM, and the role of the Galactic magnetic fields in star formation. In addition to our science, we will seek to fulfill the role of an ``observatory" balloon by opening up $30\%$ of our flight to the community as a shared-risk program to offer access to the polarized submillimeter sky that is lacking from ground-based experiments.
\bibliography{References}
\bibliographystyle{spiebib} 

\end{document}